\documentclass[twocolumn,preprintnumbers,amsmath,amssymb,prl ]{revtex4-1}

\usepackage{times}

\usepackage{amsmath}
\usepackage{amsfonts}
\usepackage{amssymb}
\usepackage{graphicx}
\usepackage{color}

\newcommand*{\citen}{}
\DeclareRobustCommand*{\citen}[1]{%
  \begingroup
    \romannumeral-`\x 
    \setcitestyle{numbers}%
    \cite{#1}%
 \endgroup
}

\definecolor{mypink}{RGB}{219, 48, 122}
\definecolor{mygreen}{RGB}{51, 153, 102}
\definecolor{brown}{RGB}{165, 42, 42}

\newcommand{\hn}{$\mathrm{h\nu}$}

\newcommand{\ZU}{$\mathrm{Z}$-$\mathrm{U}$}
\newcommand{\ZA}{$\mathrm{Z}$-$\mathrm{A}$}

 \newcommand{\AZA}{$\mathrm{A}$-$\mathrm{Z}$-$\mathrm{A}$}

\newcommand{\aGeTe}{$\alpha$\nh GeTe}
\newcommand{\aaGeTe}{$\alpha$\nh GeTe}
\newcommand{\GMT}{Ge$_{1-x}$Mn$_x$Te}
\newcommand{\GMTD}{Ge$_{0.87}$Mn$_{0.13}$Te}

\newcommand{\n}[1]{$n$\nobreakdash-\hspace{0pt}}
\newcommand\nh{\mbox{-}}
\newcommand\degC{$\,^{\circ}\mathrm{C}$}
\newcommand\dg{$^{\circ}$}

\newcommand{\Pz}{$P_{z}$}
\newcommand{\Pxy}{$P_{x,y}$}
\newcommand{\Px}{$P_{x}$}
\newcommand{\Py}{$P_{y}$}
\newcommand{\ooo}{$\langle 111 \rangle$}

\begin{document} 

\title {Operando imaging of all-electric spin texture manipulation in ferroelectric and multiferroic Rashba semiconductors}

\author{J.~Krempask{\'y}${}^{1}$, S.~Muff${}^{1,2}$, J.~Min{\'a}r${}^{3}$ , N.~Pilet${}^{1}$, M.~Fanciulli${}^{1,2}$, A.P.~Weber${}^{1,2}$,\\ V.V.~Volobuiev${}^{4,8}$, M.~Gmitra${}^{5}$,  C.A.F. Vaz${}^{1}$, V.~Scagnoli${}^{6,7}$,   G.~Springholz${}^{8}$, J.~H.~Dil${}^{1,2}$
}

\affiliation{
$^1$Swiss Light Source, Paul Scherrer Institut, CH-5232 Villigen PSI, Switzerland\\
$^2$Institute of Physics, \'Ecole Polytechnique F\'ed\'erale de Lausanne, CH-$1015$ Lausanne, Switzerland\\
$^3$New Technologies-Research Center University of West Bohemia, Plze{\v n}, Czech Republic\\
$^4$National Technical University, Kharkiv Polytechnic Institute, Frunze Str. 21, 61002 Kharkiv, Ukraine\\
$^5$Institute for Theoretical Physics, University of Regensburg, 93040 Regensburg, Germany\\
$^6$Laboratory for Mesoscopic Systems, Department of Materials, ETH Zurich, 8093 Zurich, Switzerland\\
$^7$Laboratory for Multiscale Materials Experiments, Paul Scherrer Institute, 5232 Villigen PSI, Switzerland\\
$^8$Institut f{\"u}r Halbleiter-und Festk{\"o}rperphysik, Johannes Kepler Universit{\"a}t, A-4040 Linz, Austria\\
}

\begin{abstract}

The control of the electron spin by external means is a key issue for spintronic devices. Using spin- and angle-resolved photoemission spectroscopy (SARPES) with three-dimensional spin detection, we demonstrate  \textit{operando} electrostatic spin manipulation in ferroelectric \aGeTe\ and multiferroic \GMT. We not only demonstrate for the first time electrostatic spin manipulation in Rashba semiconductors due to ferroelectric polarization reversal, but are also able to follow the switching pathway in detail, and show a gain of the Rashba-splitting strength under external fields. In multiferroic \GMT\ \textit{operando} SARPES reveals switching of the perpendicular spin component due to electric field induced magnetization reversal. This provides firm evidence of effective multiferroic coupling which opens up magnetoelectric functionality with a multitude of spin-switching paths in which the magnetic and electric order parameters are coupled through ferroelastic relaxation paths. This work thus  provides a new type of magnetoelectric switching entangled with Rashba-Zeeman splitting in a multiferroic system.

\end{abstract}

\maketitle

\section{INTRODUCTION} 
\begin{figure}
\centering
\includegraphics[width=8.6cm]{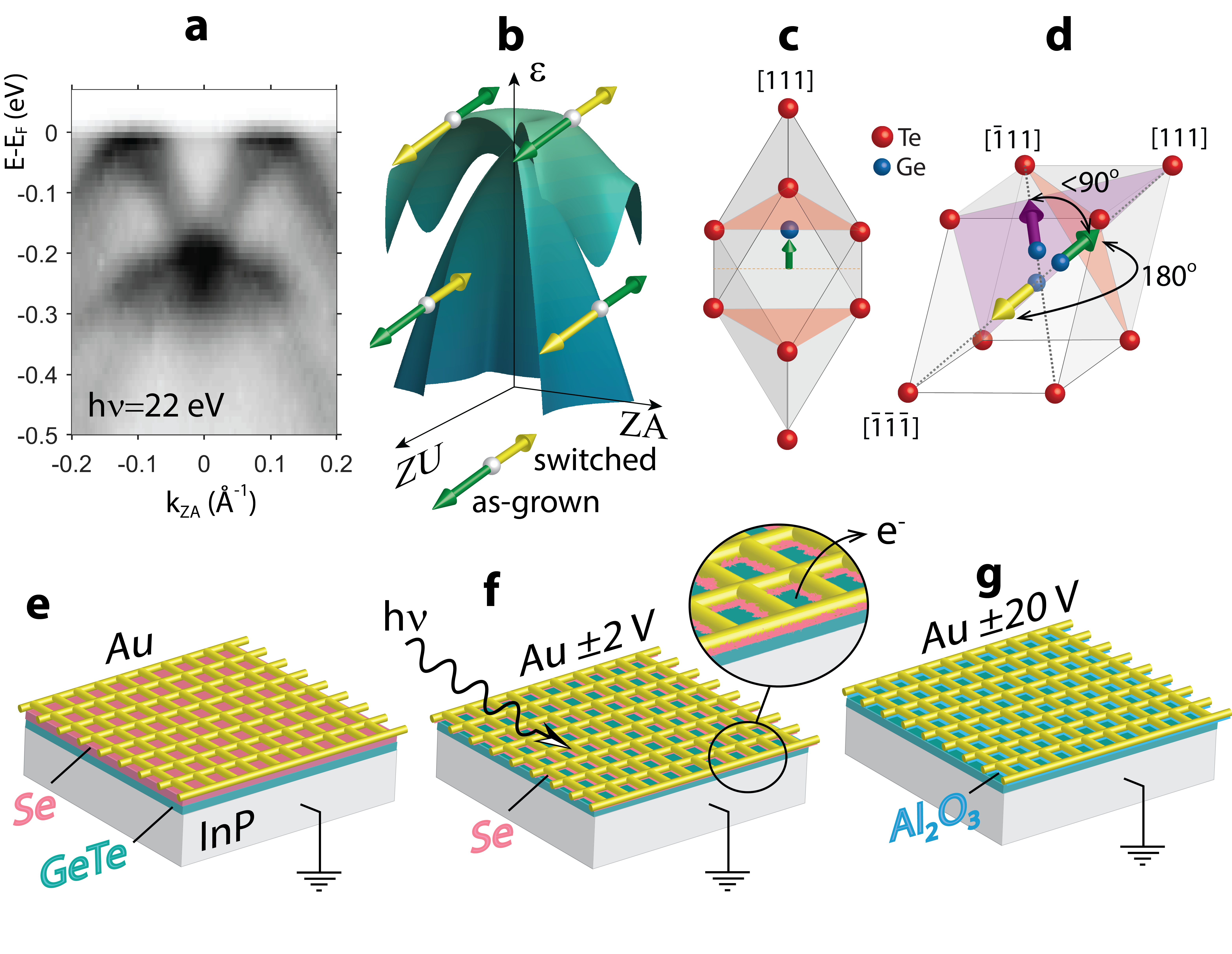}
\caption{(a) \aGeTe\ (111) ARPES band map along \AZA. (b) Schematic representation of the two upper occupied Rashba-split valence bands around the Z-point of the Brillouin zone (as-grown spin winding in green, switched in yellow). (c) \aaGeTe\ with FE displacement of the Ge atoms along [111] (green arrow). (d) Ferroelectric switching of \aaGeTe\ into the [$\bar{1}$11] oblique domain (purple arrow), and full switching into the [$\bar{1}\bar{1}\bar{1}$] direction (yellow arrow). (e) Schematic layout of the poling device used for \textit{operando} SARPES gate control before and (f) after annealing; (g) improved device with $\mathrm{Al}_2\mathrm{O}_3$ dielectric coating of the gold grid for higher bias control.}
\label{F1}
\end{figure}

Manipulating the spin texture of ferroelectrics (FE) through electric fields and of multiferroics through both magnetic and/or electric fields is a key requirement for programmable spintronic devices \cite{Maekawa_book}.   The recent discovery of giant Rashba splitting in ferroelectric \aaGeTe\ provides a promising candidate for such devices and raises the question of whether the spin texture in such a material can be modulated by electric fields by controlling the inversion asymmetry and thus also the Rashba splitting\cite{Picozzi_AdvM,Picozzi_Front}. Such devices would strongly benefit from the highest bulk Rashba spin splitting of \aaGeTe\ at room temperature\cite{JK_PRB}. This spin splitting is exemplified in Fig.~\ref{F1}a that shows the \aaGeTe\ Rashba bands near the Z-point of the Brillouin zone where the Rashba-splitting is most pronounced \cite{JK_GMT,Liebmann_GeTe,Schoenhense_GeTe}. Ideally, electric field induced ferroelectric switching  will change the spin orientation of these band as sketched in Fig.~\ref{F1}b, thus allowing a reprogramming of spin currents by external means.

\aaGeTe\ is the simplest known binary ferroelectric semiconductor with a narrow band gap\cite{Pawley_1966, JK_PRB}. Below $T_C\approx$700 K it assumes a non-centrosymmetric rhombohedral structure in which an electric dipole is formed due to a relative Ge/Te sublattice displacement along the [111] direction (see Fig.~\ref{F1}c). For moderate Mn-dopings, \GMT\ becomes ferromagnetic as well \cite{Springholz_PRL,Kriegner_2016}, which opens up new spin-based functionalities because this multiferroicity entangles the Rashba and Zeeman effects within a three-dimensional system \cite{JK_GMT}. In this new class of MUltiFErroic Rashba Semiconductors (MUFERS), we already showed that external magnetic fields switch the bulk spin texture\cite{JK_GMT}. In this work, we demonstrate that spin manipulation is also possible by \textit{electric} fields, meaning that the magnetoelectric coupling so far induced only by magnetic fields\cite{Springholz_PRL, JK_GMT} expands even to all-electric control of magnetism. Thus, from an application view, \GMT\ fulfils all criteria for mutual control of magnetism and ferroelectricity via magnetoelectric coupling effects \cite{Eerenstein2006,Fiebig2005}.   

Our epitaxial films grown by molecular beam epitaxy \cite{Springholz_PRL} on conducting InP (111) substrates were studied by means of spin- and angle-resolved photoemission under applied voltage (\textit{operando} SARPES) at 20 K, which was complemented by piezoresponse force microscopy (PFM) investigations (see Appendix A and Ref.\citen{GeTe_Gruverman,Calarco_nano,Liebmann_GeTe,JK_PRB}). To understand the functional electronic properties of \aaGeTe\ we need to consider its FE lattice structure in which the Ge atoms are displaced along [111] with respect to the six neighboring Te-atoms as sketched in Fig.~\ref{F1}c. The displacement, indicated by the green arrow, is very large reaching a value almost 10\% of the unit cell \cite{Kriegner_2016,JK_PRB}. 
An important issue for switching of the  polarization is that, due to the four-fold degeneracy of the rhombohedral lattice distortion, the ferroelectricity may form in eight different domains with individual polarization vectors pointing along different \ooo\ directions. This multi-domain structure is independent of the substrate epitaxial registry because thin films grown on InP, BaF$_2$ or Si(111) develop the same major and secondary FE-domains\cite{Springholz_PRL,JK_PRB,Lechner_2010,Calarco_JCG}. In such a multidomain structure, polarization reversal may involve intermediate steps via oblique [$\bar{1}$11] domains (purple arrow in Fig.~\ref{F1}d), typically leading to a whole realm of FE-fatigue effects \cite{Lupascu_book}.

\section{Experimental methods and setups}
SARPES allows one to directly determine the spin texture of the electronic band structure and therefore probe \textit{in operando} the reversal of the Rashba spin-polarization as a function of the applied electric field. The layout of the novel device structure developed for such measurements is shown in Fig.~\ref{F1}e and consists of a protective Se-cap on top of the \aaGeTe, respectively, \GMT\ epilayers onto which a Au-mesh is placed as a top gate (see Appendix A). Two strategies were employed in the device assembly. In the first (Fig.~\ref{F1}f), the \textit{in situ} desorbed Se-cap sticks to the bottom-side of the Au-mesh providing a quasi-insulating contact between the \hbox{Au-gate} and the semiconducting \aaGeTe. The second assembly (Fig.~\ref{F1}g) employs an Al$_2$O$_3$ layer predeposited on the Au mesh to provide better dielectric insulation. Consequently, higher bias voltages can be applied and the setup can withstand several annealing treatments to rejuvenate the sample after electric cycling. Both setups have similar capacitor-like structures and yield consistent results. 

\begin{figure}[h!]
\includegraphics[width=8.6cm]{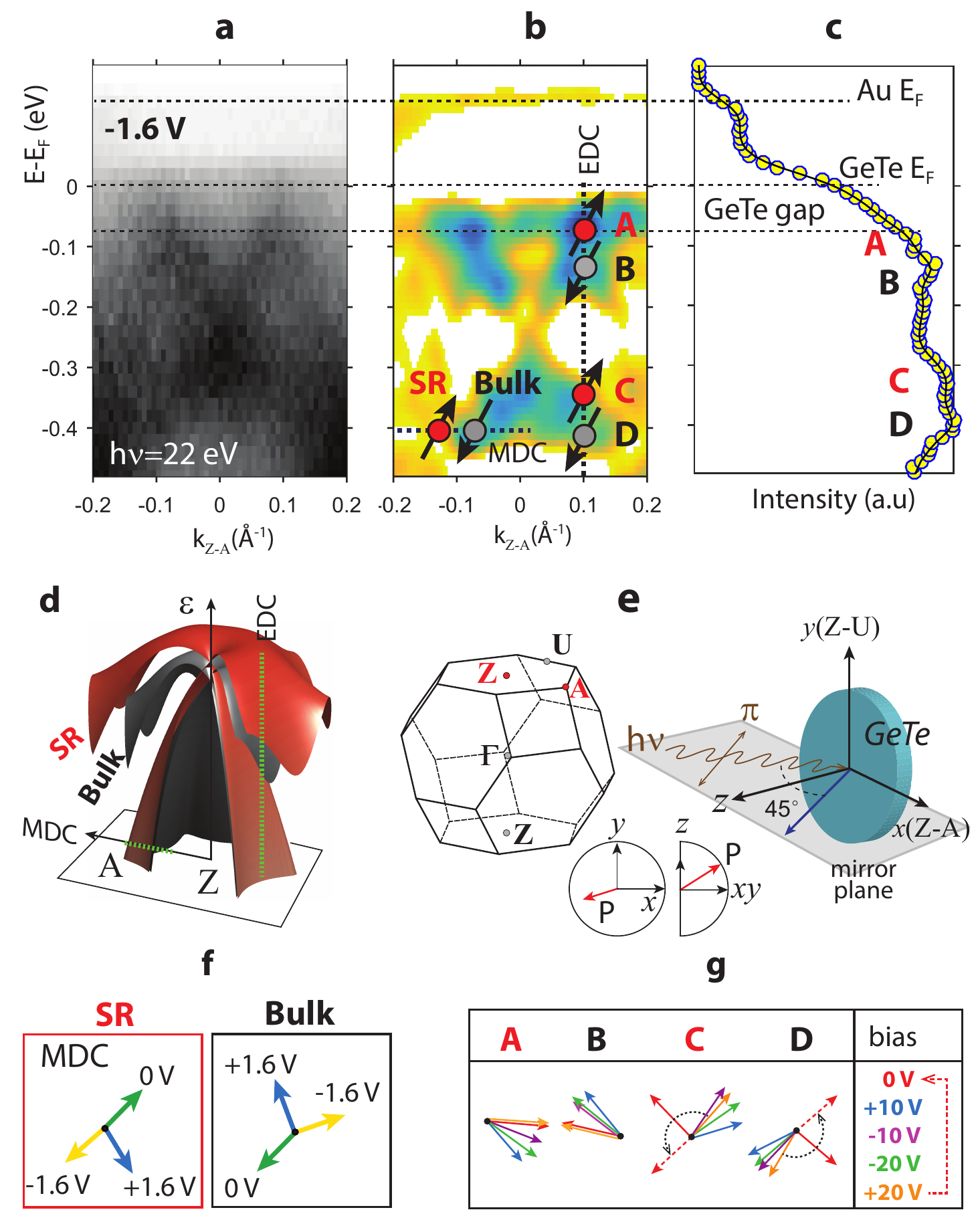}
\caption{(a) ARPES band map of \aaGeTe\ and (b) its second derivative, measured along the Z-A direction with \hbox{-1.6 V} applied gate bias. (c) EDC-cut at the location of the dashed line in (b). Indicated is the narrow band-gap above the valence band maximum and two Fermi levels from the \aaGeTe\ and the \hbox{Au-gate}. (d) A sketch of the surface electronic structure, bulk bands in black, surface-resonance band (SR) in red. (h) \aGeTe\ Brillouin zone (BZ) and experimental geometry with definition of the local \{xyz\} axes and polar representation of the 3D spin vectors. (f,g) Bias dependent in-plane \Pxy\ vectorial spin analysis of the bands indicated in (b). The bias control sequence is color-indicated, dashed arrows in (j) indicate \hbox{+20$\rightarrow$0 V} spin relaxation transition (see text).
} 
\label{F2}
\end{figure}

\subsection{Electric control of spin orientation} 
Figure \ref{F2}a-c shows ARPES spectra measured from our device with an applied bias voltage of \hbox{-1.6 V}. In agreement with our earlier studies on bare \aaGeTe\cite{JK_PRB}, near the Z-point (\hn=22 eV) we discern the valence band maximum with the giant Rashba splitting beneath a narrow gap ($\approx$90 meV). Compared to Fig.~\ref{F1}a, the quality of the spectra is affected by the top \hbox{Au-gate} electrode, yet the data is clear enough to locate the bulk bands and their surface-resonance replica (SR)\cite{JK_PRB, JK_GMT}, sketched in black and red in Fig.~\ref{F2}d. 

\begin{figure}[th!]
\includegraphics[width=8.7cm]{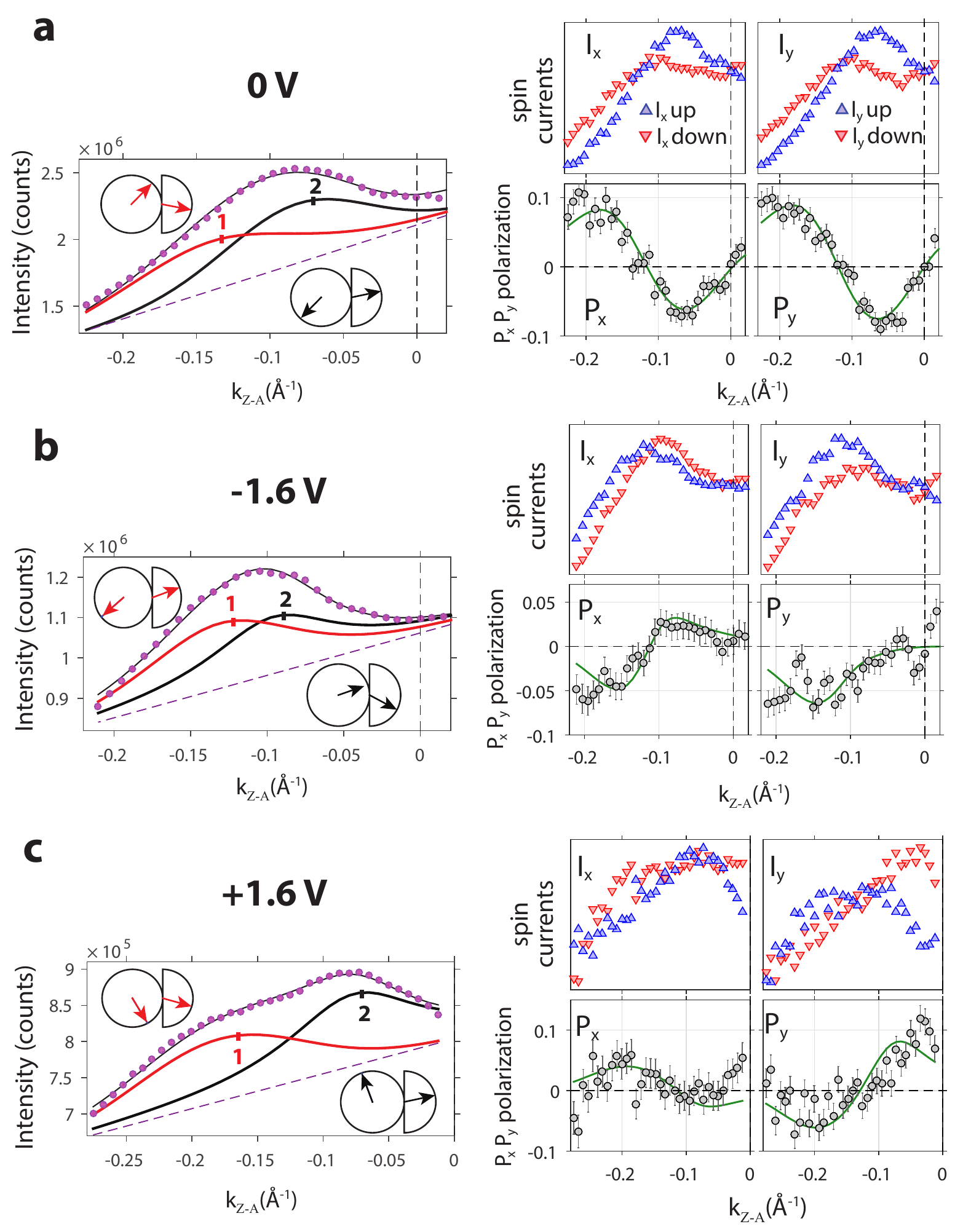}
\caption{Bias dependent in-plane \Pxy\ vectorial spins analysis from: (a) as-grown sample; (b) -1.6 V and (c) +1.6 V bias voltage. (left panels) SARPES total intensity (purple dots ) and peak-fitting of the Rashba-split bands 1 (red lines) and 2 (black lines); (right panels) spin currents $I_x$, $I_y$ and $P_x$, $P_y$ polarizations including their fits (green lines). The arrows in the left panels are the 3D spin expectation values measured within the experimental geometry depicted in Fig.~\ref{F2}e. For simplicity, in the right panels only the in-plane spin reorientations are shown. Data are obtained for the first poling device (Fig.~\ref{F1}f) along the MDC path indicated in Fig.~\ref{F2}b.} 
\label{F3}
\end{figure}

In the following, we focus on the effect of the electric field on the Rashba states in spin-detection. For this purpose, we visualize the SARPES data as spin-resolved momentum (MDC) and energy (EDC) distribution curves recorded at selected bias voltages, measured along the cuts indicated by the dashed lines in Fig.~\ref{F2}b within an experimental geometry sketched in panel (e). For clarity, panels f-g summarize the \hbox{in-plane} \textit{operando} modulation of the spin orientations. In accordance with the theoretical model in Refs.~\citen{JK_PRB,Schoenhense_GeTe}, the validity of the spin
vectors is confirmed by the fact that the bulk bands and their resonance replica show always an almost perfect anti-parallel spin arrangement. The SARPES vectorial spin fitting (see Appendix B) is detailed Figs.~\ref{F3} and \ref{F4}, respectively. 

SARPES data from first device type in Fig.~\ref{F3} identify the SR and bulk bands as peaks 1 and 2, respectively. 
As summarized in Fig.~\ref{F2}f, both show an almost perfect spin reversal in the \hbox{0$\rightarrow$-1.6 V} transition, and an incomplete spin rotation in the -1.6$\rightarrow$+1.6 V transition. This clearly demonstrates that the spin texture can be electrically manipulated by the bias applied. 

Similar electric spin manipulation is observed for the second type of device at higher bias voltages summarized in Fig.~\ref{F2}g, and detailed in Fig.~\ref{F4}. The data again shows the Rashba splitting of the bulk bands B-D and their surface resonance replica A-C with anti-parallel spins \cite{JK_PRB, Schoenhense_GeTe}. For clarity the intersection of the band A-D with the measured EDC path is sketched in Fig.~\ref{F4}d, Fig.~\ref{F4}e summarizes the spin vectorial fitting for all the bands including the \Pz\ components.

The spin rotation of bulk bands B-D appear to not fully follow the changes of applied electric poling field, presumably because each bulk band is screened by its resonance replica. This means that electric spin manipulation becomes blocked after the second cycling, an effect referred to as fatigue.

\begin{figure*}[ht!]
\includegraphics[width=\textwidth]{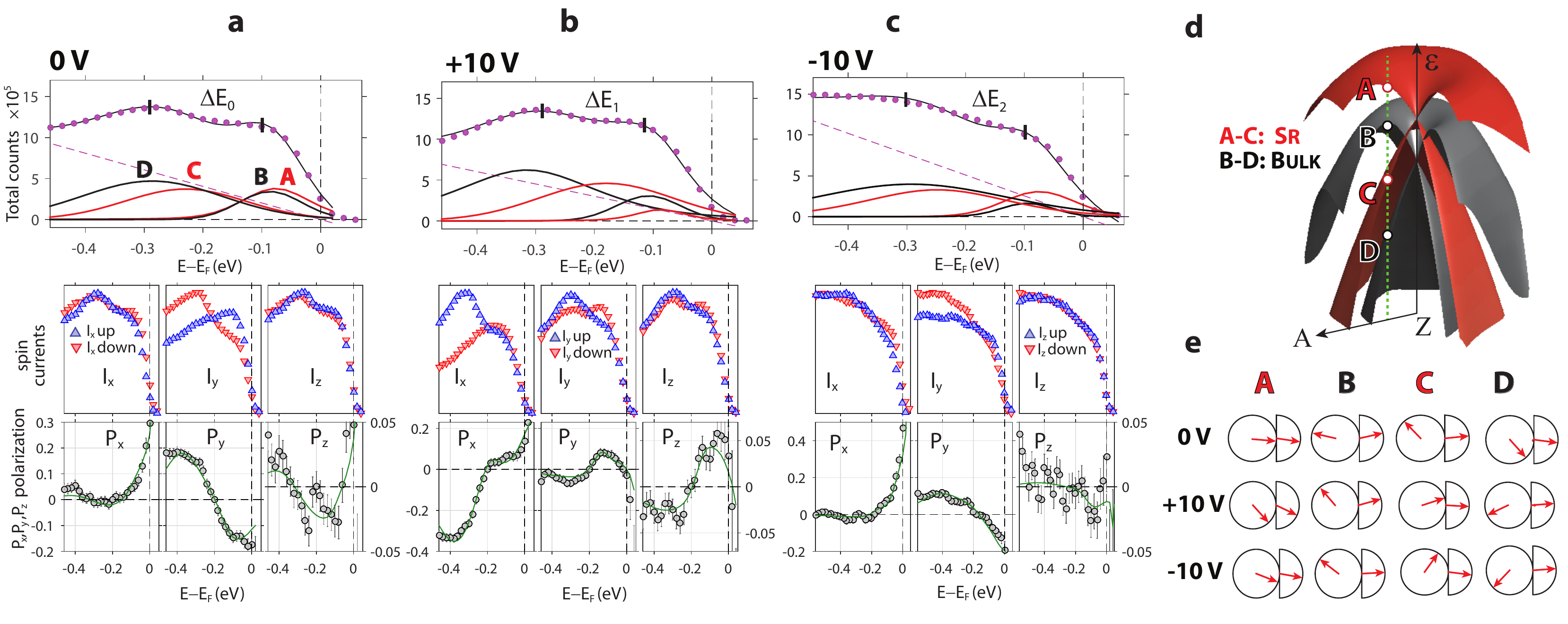}
\caption{Three-dimensional bias-dependent vectorial spins analysis from: (a) as-grown sample; (b) +10 V and (c) \hbox{-10 V} bias voltage.  SARPES total intensity $I_t$ (purple dots) and peak-fitting of the SR bands A-C (red line) and bulk bands B-D (black line) are displayed in top panels. Bottom panels display the spin currents $I_{x,y,z}$ and $P_{x,y,z}$ polarizations including their fits (green lines). (d) Schematic representation of the bulk (black) and SR bands (red). The EDC path indicated in green is intersecting the four bands A-D. (e) 3D spin vectors for each of the bands, obtained from the spin fitting for each bias voltage. Data were obtained for the second poling device sketched in Fig.~\ref{F1}g.
} 
\label{F4}
\end{figure*}

However, besides this unipolar FE-fatigue also domain wall pinning should be considered. It typically occurs when the system relaxes from a relatively high bias voltage in which the back-switching is activated by charge agglomerates at the domain walls\cite{Peng_2013}. We find that this is exactly the case for the sudden, almost 180\dg\ spin rotation in the +20$\rightarrow$0 V relaxation transition denoted by dashed arrows in Fig.~\ref{F2}g, thus demonstrating that also ferroelastic effects have to be considered in our \textit{operando} SARPES measurements. 
We note that the observed ferroelastic effects are probably related with the (meta\nh valent) resonant-bonding mechanism of GeTe\cite{Wuttig2008,GeTe_Gruverman} which leads to a decrease of the relative stability of the FE-order.

\subsection{FE domains probed by PFM and \textit{operando} SARPES}

\begin{figure}
\includegraphics[width=8.6cm]{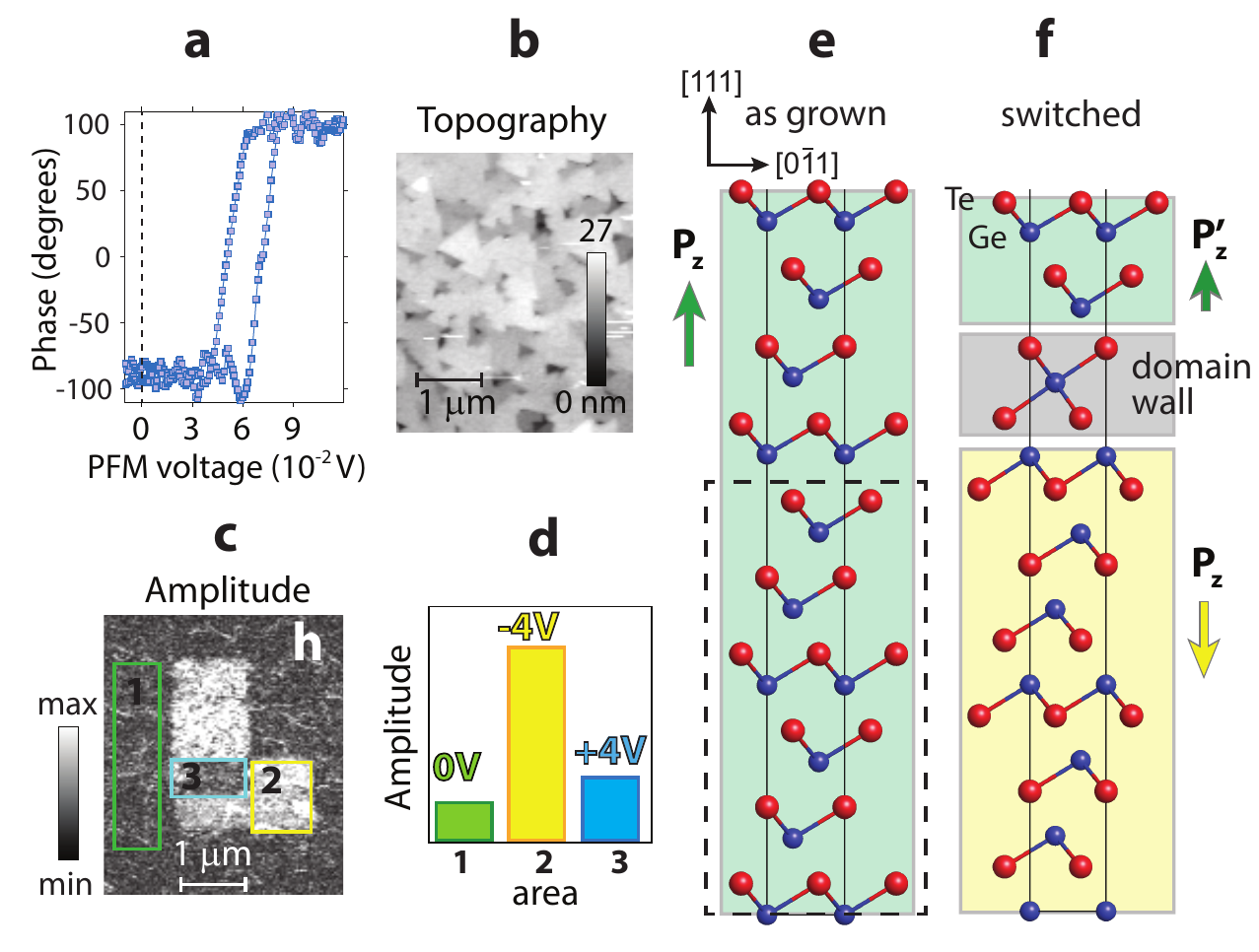}
\caption{(a) PFM vertical phase hysteresis loop measured on the  \aaGeTe/InP(111) surface. (b) Surface topography measured in atomic force microscopy mode. (c) PFM domain mapping and (d) integrated signal from selected areas (pristine area 1 in green, first written domain 2 in yellow, next overwritten domain 3 in blue). (e) As-grown unit cell of surface relaxed \aaGeTe\ with out-of-plane FE polarization (green arrow). (f) Surface relaxation after switching the \aaGeTe\ bulk atoms in the dashed rectangle indicated in (e).}
\label{F5}
\end{figure}

To elucidate the polarization reversal induced by applied electric fields and the reduced spin rotation after the initial poling cycle, PFM studies  were performed on the samples as shown in Fig.~\ref{F5}a-d. 
In this technique, by excitation of the sample via an AC voltage applied between surface and probe tip the piezoelectric effect induces an oscillation of the cantilever deflection. Its amplitude is a measure for the absolute magnitude of the FE polarization, and its phase corresponds to the polarization direction. Fig.~\ref{F5}a shows the hysteresis loop of the PFM phase as a function of DC tip voltage, evidencing the reversal of the polarization direction in the \aaGeTe\ films induced by the electric field, which corresponds exactly to the spin-reversal that we observe by SARPES. However, the shift of the hysteresis loop, similar to that in Ref.\citen{Liebmann_GeTe}, suggests a preferential built-in polarization direction related to the surface termination. 
Such experimental observation calls for a more accurate investigations of the PFM data, as well as theoretical models reflecting the FE-domain switching in the \aGeTe\ surface region.

Local FE poling of \aaGeTe\ by applied DC voltages is demonstrated by Fig.~\ref{F5}c-d, which presents the PFM amplitude image of differently poled regions written in the sequence 1-3 from \hbox{0$\rightarrow$+4 V$\rightarrow$-4 V}, indicated by the green, yellow and blue boxes, respectively. The corresponding integrated PFM amplitudes depicted in Fig.~\ref{F5}d reveal a significantly reduced PFM amplitude signal in the second poling cycle displayed in blue, similar to what we observe in SARPES by measuring only partial spin reorientation after the second poling. This suggests the existence of depolarization fields due to unipolar FE-fatigue near domains and/or grain boundaries by charge agglomerates\cite{Genenko_2015, Lupascu_PZT_fatigue}. The characteristic \aaGeTe\ thin film topography (Fig.~\ref{F5}b) originating from stacking faults\cite{Lechner_2010} are primary candidates for such carrier agglomeration, also reflected in the \aGeTe\ intrinsic \hbox{$p$-type} transport properties \cite{Edwards_PRB}. Charge agglomeration is expected to occur in the sub-surface region and screens the bulk bands, hence hindering the full spin reversal in repetitive electric cycles in agreement with our \textit{operando} SARPES results.
Based on the fact that PFM-phase switching seen in Fig.~\ref{F5}a remains unchanged after several hours also in domain mapping\cite{JK_PRB}, we rule out charging artifacts induced by PFM in the FE-switching.

Due to the small escape depth of photoelectrons,  SARPES is a surface sensitive technique and therefore the role of film surface must be taken into account. To this end, we performed \textit{ab initio} calculations based on density functional theory (see Appendix C) to assess the equilibrium surface structure as a function of applied electric field. In the calculations, the bulk FE switching was implemented by switching the inner electric field in the bulk region (dashed rectangle in Fig.~\ref{F5}e), which results in a substantial structural rearrangement of the surface region as shown by Fig.~\ref{F5}f. Given that the \hbox{Te-terminated} surface is preferred for \aaGeTe\cite{JK_PRB,Liebmann_GeTe, Deringer_JPC}, the strong electronegativity of the surface Te-atoms imposes an unipolar FE-polarization at the surface. 

The surface relaxation associated with FE switching induces a sub-surface FE domain wall (similar to the rock salt \hbox{$\beta$-GeTe} crystal structure), underneath of which the bulk \aaGeTe\ FE-order is reversed (yellow area in Fig.~\ref{F5}f). 
This triple-layer domain wall separating oppositely oriented domains is an essential ingredient in the FE-switching mechanism, which involves swapping between the longer and shorter Ge-Te bonds\cite{GeTe_Gruverman}. Near the surface the formation of the domain wall gives rise to a depolarization field $P{_z}'$ and asymmetric hysteresis loops such as seen in Fig.~\ref{F5}a. 
A way to test the contribution of unipolar FE-fatigue to the switching behavior is thermal annealing below the FE Curie point\cite{Lupascu_PZT_fatigue}. Indeed upon re-annealing of the samples to temperatures around 250-280\degC\, the spin rotation inside the first cycle observed by SARPES is restored to the initial value of the as grown samples. This is consistent with unipolar FE-fatigue due to charge agglomeration at \aaGeTe\ grain boundaries and support the theoretical model depicted in Fig.~\ref{F5}f, as well as the asymmetric PFM hysteresis loop observed in Fig.~\ref{F5}a.

\begin{figure*}[ht!]
\includegraphics[width=0.7\textwidth]{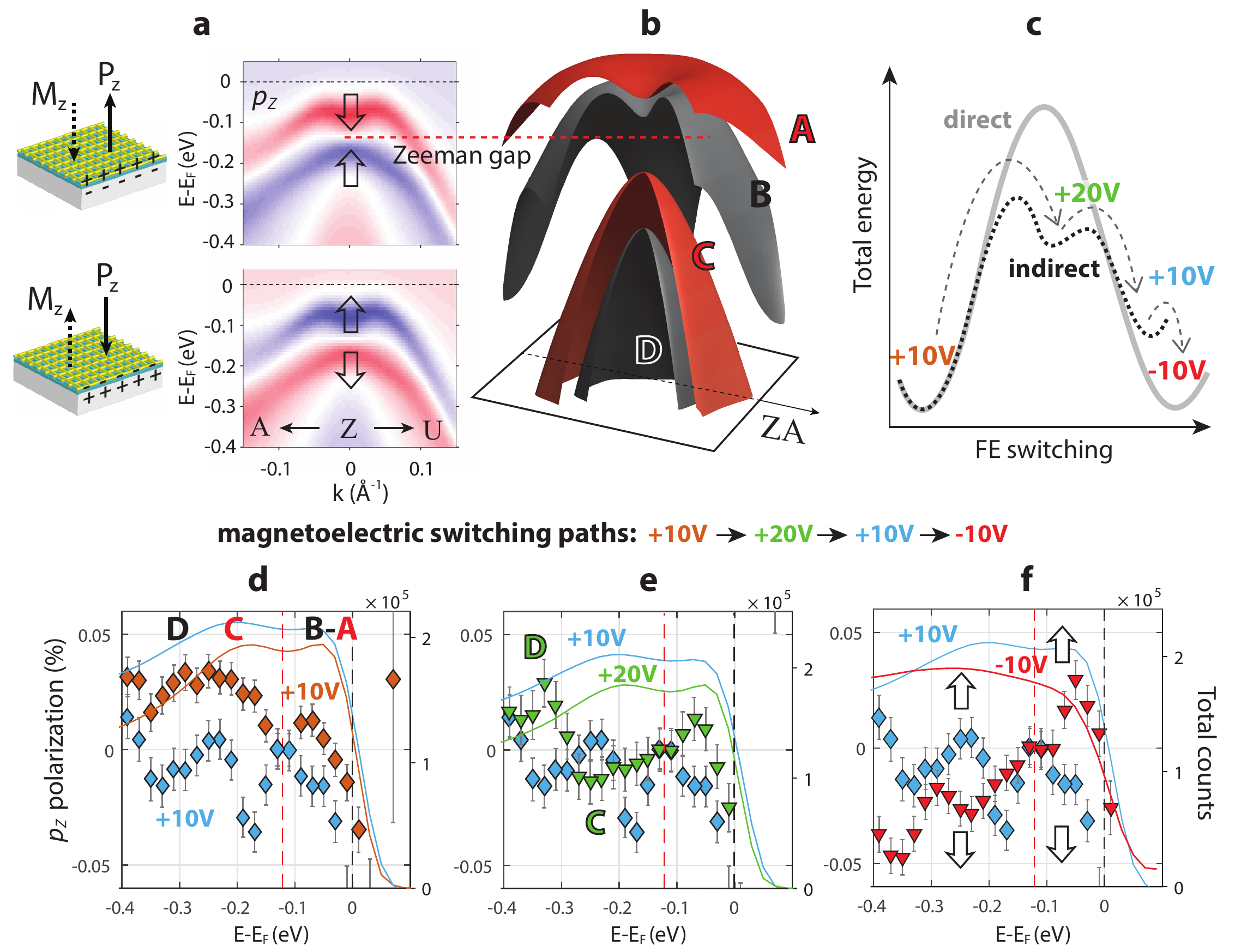}
\caption{(a) Calculated magnetoelectric coupling of the out-of-plane spin polarization $p_z$ for the bulk bands of \GMTD\ (from Ref.\citen{JK_GMT}); (top) as-grown configuration, (bottom) switched configuration. (b) Schematic representation of the \GMT\ surface electronic structure with Rashba-Zeeman split bulk bands B-D and their surface resonance replica A-C. (c) Schematic diagram of direct and kinetically favored indirect multiferroic switching paths. (d,e,f) SARPES data resolved across the Zeeman gap in normal emission (left axis, symbols), overlaid with spin-integrated EDC intensities (right axis, full lines). The arrows in panel (f) indicate the $p_z(E)$ dependence switching upon bias reversal of the gate.
}   
\label{F6}
\end{figure*}

\subsection{Control of Rashba coupling constant by electric fields} 
Inspection of the spin-fits of Fig.~\ref{F4}a-c reveals significant changes in the energy splittings $\Delta\mathrm{E_{1,2,3}}$ of the bulk Rashba bands when the samples are poled by the applied electric fields. Because the spin-fits of the bulk bands B-D (black lines in Fig.~\ref{F4}a-c) essentially coincide with the total EDC intensity in Fig.~\ref{F2}c, we can evaluate these changes from the EDCs (purple dots in Fig.~\ref{F4}a-c). Compared with the measurements of the as-grown films we find that the $\pm$10 V bias variation changes the energy splitting of the bulk Rashba bands B-D by $\pm$8\%. This is a significant change of the Rashba parameter $\alpha_R$ due to poling of the material. The 8\% increase in $\alpha_R$ implies that the initial poling field is able to align differently oriented FE domains present in the as-grown state, thus increasing the macroscopic polarization, consistent with the increase of PFM amplitude after the first poling process (yellow bar in Fig.~\ref{F5}d). Indeed, this effect can explain the deviation between the experimental Rashba constant measured for as grown \aaGeTe\ films of $\approx$4.2 eV\AA\cite{JK_PRB}, and the theoretically predicted value of $ \approx$5 eV\AA\cite{Picozzi_AdvM}.

\section{Electric field-induced magnetization reversal in G\MakeLowercase{e$_{1-\mathrm{\MakeLowercase{x}}}$}M\MakeLowercase{n$_\mathrm{\MakeLowercase{x}}$}T\MakeLowercase{e}.} 
We now turn to electric spin manipulation of multiferroic \GMT. In comparison to ferroelectrics, in MUFERs there is an additional spontaneous magnetization superimposed over the FE polarization\cite{JK_GMT}. Even more, because of the magnetoelectric coupling between magnetization and polarization, field-induced switching of polarization can also induce a switching of magnetization, thus providing additional means of spin control. A particularly interesting feature of \GMT\ is that the easy axis of magnetization ($M_z$) and the electric dipole moment ($P_z$) are colinear and perpendicular to the surface \cite{Springholz_PRL, JK_GMT}. As a result, the Rashba bands are Zeeman-split and thereby assume an additional out-of-plane spin polarization $p_z$ around the Z-point\cite{JK_GMT}. As illustrated by Fig.~\ref{F6}a and verified in Ref.\citen{JK_GMT}, this out-of-plane spin component $p_z$ is directly linked to the magnetization direction. Thus, the switching of $M_z$ can be directly detected by SARPES. 

Figure \ref{F6}d-f presents the measured $p_z$ polarization of the \GMTD\ device across the Zeeman gap in normal emission under different bias conditions. Clearly, the characteristic $p_z(E)$ dependence switches sign upon bias reversal of the gate, indicating an induced reversal of the magnetization by the applied electric field. This experimental result presents the first unambiguous and direct evidence for the existence of a strong magnetoelectric coupling between the electric dipoles and the magnetic moments in this material.  

Contrary to \aaGeTe, where the spin gate control becomes blocked after the first poling cycle, SARPES of \GMTD\ is found to change for each change of the external electric field. This is demonstrated by Fig.~\ref{F6}d-f, where the $p_z$ data, collected at each step of the colour-coded electric field sequence, is presented in comparison with that recorded at +10V (blue symbols). Evidently, each electric field change induces a significant change of the spin polarization, indicating a significantly weaker pinning of the FE polarization compared to \aaGeTe. This is explained by the fact that the lattice distortion and off-center displacement of the Ge atom with respect to the Te atoms in \GMTD\ decreases with increasing Mn content\cite{Springholz_PRL,Kriegner_2016}. This evidently reduces the energy barriers for switching of the atomic positions in the FE reorientation and thus leads to a softening of the FE properties. As illustrated in Fig.~\ref{F6}c, moreover, the coupling to the magnetization can give rise to complex switching paths that will result in unconventional spin texture evolutions as a function of biasing sequence.

\section{Conclusing remarks} 
In summary, we have demonstrated all-electric spin manipulation in  ferroelectric and multiferroic Rashba semiconductor devices using \hbox{\textit{operando}} spin-resolved photoemission spectroscopy.  The results not only give direct evidence for tuning of the spin texture by electrostatic gates, but also reveal a sizeable increase of the Rashba coupling strength achieved by poling of the material. Moreover, by the out-of-plane spin polarization we demonstrated magnetisation reversal in multiferroic \GMTD\ by applying electric fields, which provides unambiguous evidence for the strong magnetoelectric coupling between the ferroelectric and magnetic orders in this system. Our experimental findings open up a promising path toward robust and programmable semiconductor-based spintronics with functionally coupled electronic and magnetic properties. However, ferroelectric fatigue and ferroelasticity play a crucial role in determining the device performance. Therefore, robust all-electric spin switching between remnant polarization states requires further improvements in domain stabilization and sample growth.

\section{acknowledgements}
Constructive discussions with \hbox{V. N. Strocov} are gratefully acknowledged. This work was supported by the Swiss National Science Foundation Project PP00P2\_144742 1 and the Austrian Science Funds Project SFB F2504-N17 IRON. M.G. acknowledges the German funding agency DFG SFB~689 and J.M. the CENTEM Project, Reg.No.CZ.1.05/2.1.00/03.0088, CENTEM PLUS (LO1402) and COST LD15147.

\section{Appendix} 

\subsection{A: Experimental techniques}
Experiments were performed on 200 nm thick \aGeTe\ and \GMTD\ films grown by molecular beam epitaxy on semiconducting InP(111) substrates. Samples were capped by a protective stack of amorphous Te and Se layers which were removed by annealing \textit{in-situ} just before the photoemission experiments. SARPES experiments were performed at the COPHEE end-station of the Surfaces and Interfaces Spectroscopy beamline at the Swiss Light Source (SLS) synchrotron radiation facility, Paul Scherrer Institute, Switzerland, using $p$-polarized, \hn=22 eV photons. The Omicron EA125 hemispherical energy analyzer is equipped with two orthogonally mounted classical Mott detectors\cite{Hoesch_JESRP}. The whole setup allows simultaneous measurements of all three spatial components of the spin-polarization vector for each point of the band structure. The SARPES data were measured with the sample azimuths \ZU\ or \ZA\ aligned perpendicular to the scattering plane as denoted in Fig.~\ref{F2}e. The angular and combined energy resolution were 1.5\dg~and 60 meV, respectively. In spin-integrated mode these resolutions were set to 0.5\dg~and 20 meV.
All data were collected with the sample kept at a temperature of 20 K. Temperature was measured by a Si diode placed near the sample.
The InP substrate was grounded during the SARPES measurements and a bias was applied to the Au-mesh. Care was taken to avoid a direct grounding of the front of the sample. The measured kinetic energies were adjusted according to the applied voltage.

Piezo-force microscopy was performed at the NanoXAS endstation at the SLS using a plain platinum tip at room temperature\cite{Pilet2012}. The sample topography and PFM channels (amplitude and phase) are measured simultaneously. The PFM is measured close to resonance; no cross-talking with topography is visible which validate the quality of our PFM data seen in Fig.~\ref{F5}c.

The electro-formed Au-mesh used in the poling device had a nominal aperture of 64 $\mu$m, thickness 4 $\mu$m and 70\% open area. The nominal aperture is much larger compared to domain wall size, which, in first approximation, is around 1 micron according to the PFM topography in Fig.~\ref{F5}b. As the beam spot size is ca. 100 microns, and the Au-mesh hole is much larger compared to the domain wall size, we do not expect artefacts in \textit{operando} SARPES due to domain wall motions.

\subsection{B: SARPES vectorial analysis}
By means of the two orthogonal Mott detectors of the COPHEE experimental station it is possible to simultaneously measure the two in-plane directions and the out-of-plane direction, constituting a three-dimensional (3D) spin-resolved data set. Such a typical SARPES data set as presented in the main text consists of a total intensity spectrum $I_t$ and polarization spectra along the three spatial directions $P_{x,y,z}$ described in Fig.~\ref{F2}e. From this data set the spin resolved spectra can be obtained by projecting the results on these spatial directions. The measured polarization and total intensity yield spin-resolved intensity spectra calculated as

\begin{equation}
\begin{split}
  I_{x,y,z}^{up} = (1+P_{x,y,z})I_t/2 \\
  I_{x,y,z}^{down} = (1-P_{x,y,z})I_t/2
\end{split}
\label{E1}
\end{equation}

\noindent In a well established fitting routine the photoemission spectrum is first dissected into individual peaks and background and afterwards all polarization directions are fitted simultaneously to obtain spin vectors\cite{Fabian_NJP}. Figure \ref{F3} shows how the total intensity $I_t$ of the MDC is fitted with Voigt functions and a linear background. Similarly, Figure \ref{F4} shows the vectorial fitting of the energy-distribution curve (EDC) by Voigt functions and a background for a momentum near 0.1 1/\AA\ along the \ZA\ direction indicated in Fig.~\ref{F2}b. We note that in order to asses the spin manipulations upon gate control in \aaGeTe. the full 3D spin analysis is mandatory because the \aaGeTe\ Rashba-splitting manifest a canted spin arrangement also in the in-plane spin texture\cite{JK_PRB}.

The polarization curves in Fig.~\ref{F3} (Fig.~\ref{F4}) are modeled until the best fit is reached by simultaneously fitting the MDC (EDC) intensity (purple dots) and the polarizations \Px, \Py~and \Pz\ (green lines). All the MDC SARPES fits in Fig.~\ref{F3} unambiguously identify the surface resonance as peak 1 in red and the bulk as peak 2 in black.

Concerning the \textit{operando} SARPES we emphasize that during the 0 V$\rightarrow$-1.6 V transition in Fig.~\ref{F3}a-b, i.e. during the first poling of the as-grown \aGeTe\ film, there is almost an ideal switching of all the spin polarization components, well observed also in the raw spin polarization curves $P_{x,y}$. These experimental observations are consistent with complete reversal of the spin vectors as depicted in Fig.~\ref{F1}b.

\subsection{C: First principles calculations}
To calculate structural properties associated with the ferroelectric switching in $\alpha$-GeTe, we perform density functional theory calculations employing the {Quantum ESPRESSO} package~\cite{Giannozzi2009:JPCM}.
We used projector augmented wave pseudopotentials with kinetic energy cutoff of 40~Ry for wave functions and 240~Ry for charge density. For the exchange-correlation functional we used the generalized gradient approximation \cite{Perdew1996:PRL}.
The surface properties were studied in a slab geometry. The $\alpha$-GeTe
crystal along [111] direction contains 18 atomic layers with hydrogen passivated Ge surface and physically relevant Te (111) surface. The Brillouin zone was sampled with $9\times 9$ $k$ points. Atomic positions were relaxed using the quasi-newton algorithm based on the trust radius procedure.

\bibliographystyle{apsrev4-1}
%

\end{document}